\RequirePackage{fix-cm}
\documentclass[twocolumn,epjc3]{svjour3}  
\smartqed  
\RequirePackage{graphicx}
\usepackage{multirow}
\RequirePackage{amsmath}
\journalname{Eur. Phys. J. C}
\begin{document}

\title{Effect of Strong Field Space-time Features on Vacuum Pair
}


\author{B.~An\thanksref{addr1}
        \and
        N.~S.~Lin\thanksref{e1,addr1,addr2}
        \and
        C.~K.~Li\thanksref{e2,addr3}
        \and
        M.~Jiang\thanksref{addr1,addr2}
        \and
        Y.~J.~Li\thanksref{e3,addr1,addr2}
}

\thankstext{e1}{e-mail: phy.nslin@gmail.com}
\thankstext{e2}{e-mail: chuankeli@163.com}
\thankstext{e3}{e-mail: lyj@aphy.iphy.ac.cn}


\institute{State Key Laboratory of Tunnel Engineering, China University of Mining and Technology, Beijing 100083, China\label{addr1}
          \and
          School of Science, China University of Mining and Technology, Beijing 100083, China\label{addr2}
          \and
          Laboratory of Zhongyuan Light, School of Physics, Zhengzhou University, Zhengzhou 450001, China\label{addr3}
}

\date{Received: date / Accepted: date}

\maketitle

\begin{abstract}
The relativistic dynamics of bound states across inertial reference frames are investigated using the computational quantum field theory (CQFT). The results reveal that the spatiotemporal properties of bound states within a given potential well are strictly frame-dependent. Crucially, this spatiotemporal modulation of the external field enables a reduction in the laser intensity threshold required for vacuum electron-positron pair creation. Analytical and numerical calculations demonstrate that this threshold reduction originates from the Lorentz transformation of the four-momentum, which reshapes the vacuum excitation pathways in phase space. By developing CQFT, we establish a comprehensive framework in which relativistic effects intrinsically govern the quantum vacuum decay process.
\end{abstract}

\section{Introduction}
\label{intro}
In the 1920s, the Dirac equation was put forward by P. A. M. Dirac, introducing the pivotal concept of the Dirac sea and famously predicting the existence of the positron \cite{Dirac1928}. Empirical confirmation of Dirac's prediction soon followed, achieved through C. D. Anderson's experimental observation of positrons \cite{Ande1993}. The first theoretical treatments of electron- positron pair creation from the vacuum in strong fields date back to the 1930s, with seminal works by F. Sauter \cite{Sauter1931} and by H. Euler and W. Heisenberg  \cite{Euler1936}. A pivotal advancement came in 1951, when J. Schwinger formalized the theory of pair production. By employing the quantum electrodynamics, he successfully recovered the pair creation rate and critical field strength in a constant electric field—a phenomenon now commonly known as the Schwinger effect \cite{schw1951}. The required field strength for this process to become significant is on the order of $E = 1.32 \times10^{16}~\text{V/cm}$, corresponding to an intense laser field of approximately $10^{29}~\text{W/cm}^2$. Employing the WKB approximation, Brezin and Itzykson uncovered the multi-photon pair creation mechanism in their seminal 1970 work \cite{Brezin1970}. The dynamically assisted Schwinger mechanism, which arises from the synergistic combination of Schwinger pair creation and multi-photon absorption, enables the observation of electron-positron pairs at laser intensities one to two orders of magnitude below the Schwinger critical value, achieved by superimposing a low-frequency strong field with a high-frequency weak field \cite{RS1}. The dynamically assisted Schwinger mechanism is considered a highly promising scheme for realizing the experimental observation of the Schwinger effect, as it has consequently attracted considerable attention in theoretical studies \cite{ZLL1,IAA1,SVC1,XGH1,HT1,YLR1,IAA2,Ib}. Furthermore, the question of how to maximize the yield of created particles through optimal field configurations has motivated extensive investigations into their space-time structures \cite{FH1,SSB1,CK1,LV20141,SSD1,JU1}. Meanwhile, the instantaneous rest frame method (IRFM) was developed, demonstrating that a subcritical potential well can transition into a supercritical configuration, thereby significantly lowering the laser intensity required for electron-positron pair production \cite{LV20141}. However, the variations in the energy distribution and the impact of bound-state lifetimes on particle production have not yet been thoroughly investigated. These issues will also be the focus of our future research.

Studying electron-positron pair creation in physical systems involving bound states has long been a key research focus in this field. The combined effects of oscillating and static potential wells on pair production have been investigated, which revealed a significant enhancement attributed to the transient excitation of field-induced bound states \cite{jiang2013,tang2013}, while other works illustrated that spatial asymmetry can enhance pair production in spatially asymmetric time-oscillating fields\cite{mamat2025}.
Moreover, the interplay between intense laser fields and external potential wells has become a subject of growing research interest. The scenario of electron-positron pair creation, arising from the head-on collision of intense laser pulses with X-ray photons in the presence of stationary Coulomb charges, has been investigated \cite{HF2019}. The process of electron-positron pair creation induced by laser fields in the presence of a highly charged model nucleus has been explored, with particular emphasis on how the energy of the bound states affects vacuum pair creation \cite{SU2020,SU2021}. We demonstrate that phase-controlled colliding laser pulses can probe the nucleus-induced deformation of the Dirac vacuum \cite{CK2023}. Furthermore, the bound states diving into the negative energy continuum give rise to Feshbach resonances in static supercritical potential wells \cite{LV2013,LV2014,LIU2014,LIU2015,XXZHOU2023}. In order to predict the energies and lifetimes of bound states, the complex coordinate transformation method was developed and applied~\cite{LV2013,LV2014}.

In this paper, we apply both CQFT and IRFM to investigate electron-positron pair creation induced by a relativistically moving potential well. Within the CQFT framework, we have developed a novel method for predicting bound-state properties that offers a simpler yet more effective alternative to complex coordinate transformation method. By defining the tunneling probability for electrons occupying negative-energy states, we analyze the number, energy, and lifetime of the resulting bound states. Our results reveal that these properties are not Lorentz invariant—distinct bound-state characteristics are perceived in different reference frames.

This paper is organized as follows. In Sec.~\ref{2}, we briefly review and further develop the CQFT, and introduce the external field configuration. Section~\ref{3} is dedicated to investigating electron-positron pair creation induced by a relativistically moving potential well, using the electron tunneling probability as a visible tool. In Sec.~\ref{4}, we examine the relativistic transformation of bound-state properties across different reference frames, and present a detailed analysis and discussion of the results. Finally, we summarize our findings and conclude in Sec.~\ref{5}.

\section{Theoretical Method and External Field Model}
\label{2}
\subsection{The computational quantum field theory}
\label{2a}
In this work, we employ CQFT, as it offers a direct computational framework that is particularly well suited for investigating time-dependent and spatially inhomogeneous field configurations. Within the framework of CQFT, the spacetime evolution of the electron-positron field operator $\hat{\psi}(z,t)$ is governed by the Dirac equation~\cite{Cheng2010,jw1999,Su2012,lv2018,zhou2021,SU2022,su2025}
\begin{equation}
i \frac{\partial \hat{\psi}(z,t)}{\partial t} = \left[ \sigma_1 cp + \sigma_3 c^2 + V(z,t) \right] \hat{\psi}(z,t),
\label{eq1}
\end{equation}
where $\sigma_1$ and $\sigma_3$ are the Pauli matrices and $c$ is the speed of light in vacuum.  In this paper, we use atomic units, with $e=m_e=\hbar=1$ and $c=137.036$. The external electric field is modeled through the scalar potential $V(z,t)$. Notably, homogeneity in the $x$ and $y$ directions allows the reduction of the Dirac spinor to two components~\cite{Cheng2010,jw1999,Su2012,lv2018,zhou2021,SU2022,su2025}. The field operator can be expanded in terms of the particle annihilation operator $\hat{b}_p$ and the antiparticle creation operator $\hat{d}_n^\dagger$,
\begin{equation}
\begin{split}
\hat{\psi}(z,t) &= \sum_p \hat{b}_p(t) u_p(z) + \sum_n \hat{d}_n^\dagger(t) u_n(z) \\
&= \sum_p \hat{b}_p u_p(z,t) + \sum_n \hat{d}_n^\dagger u_n(z,t).
\end{split}
\label{eq2}
\end{equation}
Here, $u_p(z,t)$ and $u_n(z,t)$ evolve from the field-free positive-energy state $u_p(z)$ and negative energy states $u_n(z)$, respectively. The time evolution of the electron annihilation operator $\hat{b}_p(t)$ and the positron creation operator $\hat{d}_n^\dagger(t)$ is given by
\begin{equation}
\begin{split}
\hat{b}_p(t) &= \sum_{p'} \hat{b}_{p'} \int dz~u_p^*(z)u_{p'}(z,t) \\
&+ \sum_{n'} \hat{d}_{n'}^\dagger \int dz~u_p^*(z)u_{n'}(z,t),
\end{split}
\label{eq3}
\end{equation}

\begin{equation}
\begin{split}
\hat{d}_n^\dagger(t) &= \sum_{p'} \hat{b}_{p'} \int dz~u_n^*(z)u_{p'}(z,t) \\
&+ \sum_{n'} \hat{d}_{n'}^\dagger \int dz~u_n^*(z)u_{n'}(z,t).
\end{split}
\label{eq4}
\end{equation}
The field operator can be decomposed into electronic and positronic components as $\hat{\psi}^{(e)}(z,t)=\sum_p\hat{b}_p(t)u_p(z)$ and $\hat{\psi}^{(p)}(z,t)=\sum_p\hat{d}_n^\dagger(t)u_n(z)$, respectively.
The spatial density of electrons is then given by the vacuum expectation value
\begin{equation}
\begin{split}
\rho^{(e)}(z,t) &= \ll \mathit{vac} \| \hat{\psi}^{(e)\dag}(z,t) \hat{\psi}^{(e)}(z,t) \| \mathit{vac} \gg \\
&=\sum_n|\sum_p\mathbf{U}_{p,n}(t)u_p(z)|^2,
\end{split}
\label{eq5}
\end{equation}
and their momentum distribution by
\begin{equation}
\rho^{(e)}(p_z,t)=\sum_{n}|\mathbf{U}_{p,n}(t)|^2.
\label{eq6}
\end{equation}

Similarly, for positrons we have
\begin{equation}
\begin{split}
\rho^{(p)}(z,t) &= \ll \mathit{vac} \| \hat{\psi}^{(p)\dag}(z,t) \hat{\psi}^{(p)}(z,t) \| \mathit{vac} \gg \\
&=\sum_p|\sum_n\mathbf{U}_{n,p}(t)u_n(z)|^2,
\end{split}
\label{eq7}
\end{equation}
and
\begin{equation}
\rho^{(p)}(p_z,t)=\sum_{p}|\mathbf{U}_{n,p}(t)|^2.
\label{eq8}
\end{equation}
The total number of created electron-positron pairs is given by the integral over the spatial density 
\begin{equation}
\begin{split}
N^{(e)}(t)
&=\int \,dz\,\rho^{(e)}(z,t)
=\sum_{p,n}\big|U_{pn}(t)\big|^2 \\
&\qquad =\sum_{p,n}\big|\langle p|U(t)|n\rangle\big|^2, \\[4pt]
N^{(p)}(t)
&=\int \,dz\,\rho^{(p)}(z,t)
=\sum_{n,p}\big|U_{np}(t)\big|^2 \\
&\qquad =\sum_{n,p}\big|\langle n|U(t)|p\rangle\big|^2,
\end{split}
\label{eq9}
\end{equation}
in the theoretical framework of vacuum pair production, the number of electrons and positrons is strictly equal, so the following relation should hold: $N^{(e)}(t) = N^{(p)}(t)$.

In our model, the matrix elements $U_{p,n}(t)$ and $U_{n,p}(t)$ of the unitary time evolution operator serve as the fundamental building blocks of the CQFT calculation. Here, $U(t) = \hat{T}\exp\left[ -i \int_0^t h(t')\, dt' \right]$ is the time evolution operator. For the numerical computation, the total evolution time from 0 to $t_{max}$ is discretized into $N_t$ intervals, with a time step $\Delta t=t_{max}/N_t$ on the order of $10^{-6}$. Following the split operator technique \cite{AD1994,GR2008}, the time evolution operator for a single step can be written as 
\begin{equation}
\begin{split}
U(t + \Delta t, t) &= \exp\left(-i\frac{V\Delta t}{2}\right) \times \exp(-ih_0\Delta t) \\
&\times \exp\left(-i\frac{V\Delta t}{2}\right) + \mathcal{O}(\Delta t^3).
\end{split}
\label{eq10}
\end{equation}
The action of this operator is implemented efficiently by employing a fast Fourier transform (FFT) to switch between the spatial and momentum representations at each stage. By applying this decomposed operator over $N_t$ consecutive time steps, we obtain the evolved states from which the spatial density, the momentum distribution and the total particle number are computed.

It is noted that $\left|U_{p,n}(t)\right|^{2}$ represents the projection of the evolved negative energy state $|n(t)\rangle$ onto the initial positive energy states $|p\rangle$ at time $t$. The energy of the positive state is given by $E_p=\sqrt{p^2c^2+c^4}$, while the energy of the negative state is $E_n=-\sqrt{n^2c^2+c^4}$. Consequently, $\left| \langle E_p|E_n(t)\rangle \right|^{2}=\left|U_{E_p,E_n}(t)\right|^{2}$ gives the probability for the creation of a particle from the negative energy state $E_n$ into the positive energy state $E_p$. In this scenario, all external fields are simultaneously turned off at a specific time $t$, and the resulting time-dependent number of created particles represents the actual physical count of particles~\cite{CCG1,CKL4}. For the multi-photon pair creation, $\left| \langle E_p|E_n(t)\rangle \right|^{2}=\left|U_{E_p,E_n}(t)\right|^{2}$ describes the probability for an electron to transition from a negative energy state $|E_n\rangle$ to a positive energy state $|E_p\rangle$ at time $t$~\cite{jiang2023,ck}. In the context of Schwinger pair creation, these matrix elements can be interpreted as the tunneling probability between negative and positive energy states at time $t$.

\subsection{The moving potential well}
\label{2b}
\begin{figure}[htbp]
\centering
\includegraphics[width=0.5\textwidth]{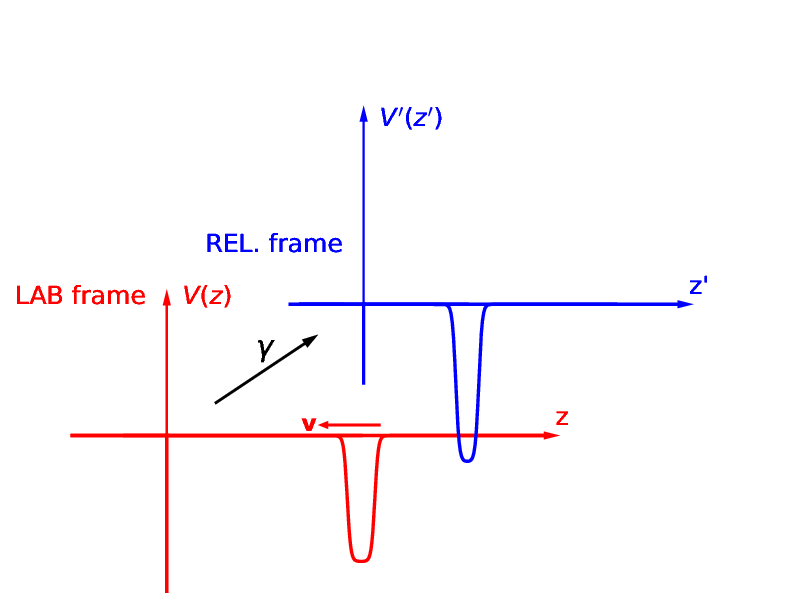}
\caption{\label{fig1} Schematic picture of the relativistic moving potential well. The potential well moves at a constant velocity along positive direction of the $z$-axis in the laboratory frame.}
\end{figure}

In this work, we study electron-positron pair creation induced by a uniformly moving potential well as shown in Fig.~\ref{fig1}. In the laboratory frame (Lab frame), the potential well is described by
\begin{equation}
V(z,t) = \frac{V_0}{2} \left[ S\left(z - \frac{D}{2} + v_0 t\right) - S\left(z + \frac{D}{2} + v_0 t\right) \right],
\label{eq11}
\end{equation}
with $S(z) = 1 + \tanh(z/W)$, where $V_0$ is the potential depth, $W$ characterizes the field width, $D$ denotes the width of potential well, $v_0$ represents moving velocity of the potential well. 
We then perform a Lorentz transformation to a new frame ($z'$, $t'$) moving with velocity $v_0$ along the $z$ direction relative to the original laboratory frame~\cite{LV20141}. The corresponding Lorentz factor is $\gamma=[1-(v/c)^2]^{-1/2}$. Under this transformation, the potential well parameters become $V^{'}=\gamma V $, $W^{'}=\gamma W $, $D^{'}=\gamma D $, while the coordinates transform as $z'=\gamma(z-vt)$ and $t' =\gamma (t - vz/c^2)$. In the relative frame (Rel. frame), the potential well takes the form
\begin{equation}
V^\prime(z^\prime) = \frac{V_0^\prime}{2} \left[ S\left(z' - \frac{D^\prime}{2}\right) - S\left(z' + \frac{D^\prime}{2}\right) \right].
\label{eq12}
\end{equation}

\section{The Electron-Positron Pair Creation Induced by the Relativistic Moving Potential Well}
\label{3}

\begin{figure}[htbp]
\centering
\includegraphics[width=0.5\textwidth]{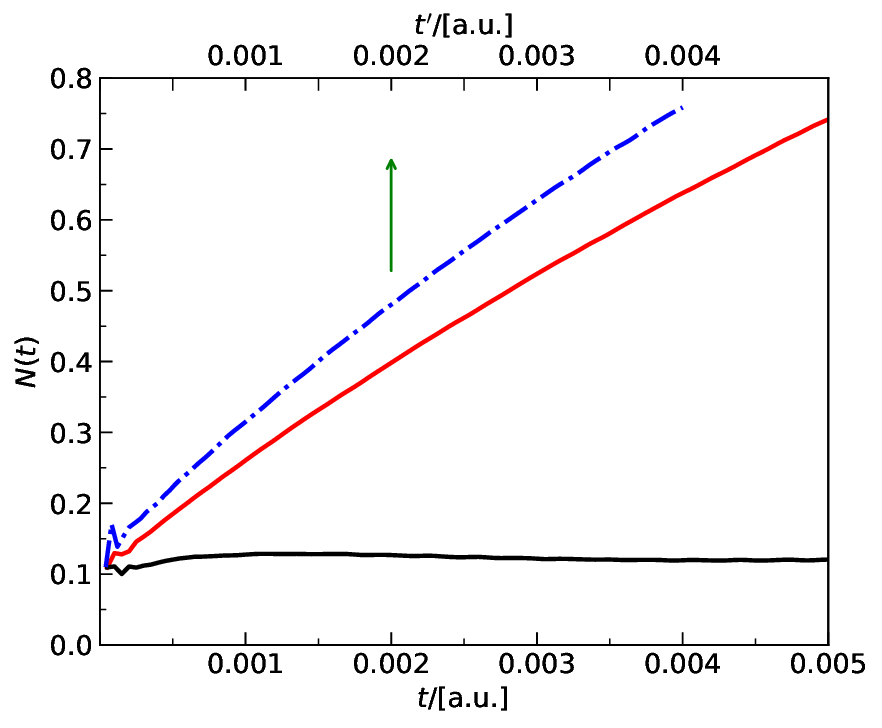}
\caption{\label{fig2} Number of created pairs as a function of time. The black solid line represents a rest potential well in the laboratory frame ($V=2.4c^2$, $W=1/c$, $D=4/c$, $v_0=0$), the red solid line represents a relativistic moving potential well in the laboratory frame ($V=2.4c^2$, $W=1/c$, $D=4/c$, $v_0=0.6c$). The blue dash-dot line represents a rest potential well in the relative frame ($V'=3c^2$, $W'=1.25/c$, $D'=5/c$, $t'=0.004$, $v^{'}=0$).}
\end{figure}

In this section, we analyze electron-positron pair creation induced by a relativistically moving potential well, considering both the laboratory frame and the relative frame. Fig.~\ref{fig2} illustrates the time evolution of the number of created particles. 
We observe that electron-positron pairs are created from the vacuum at the initial time. In our simulation, this is achieved by instantaneously turning on the potential well, which leads to vacuum decay and the creation of a small number of electron-positron pairs. The yield of these particles is negligible within our model~\cite{su2004,su2005}. To ensure that the particle creation arises from the relativistic motion of the potential well, we consider a potential well whose lowest bound state is not in resonance with the negative-energy continuum. When the potential well is at rest ($v_0 = 0$), no electron-positron pairs are created from the vacuum as shown by the black solid line in Fig.~\ref{fig2}. In contrast, electron-positron pairs are created for a relativistically moving potential well, yielding $N(t = 0.005) = 0.7418$ for $v_0 = 0.6c$ (red solid line in Fig.~\ref{fig2}). 
Furthermore, we employ the IRFM by transforming from the laboratory frame to the relative frame (where the potential well is at rest) using a Lorentz transformation. In the new frame, the number of created particles at the corresponding time is $0.7586$ (blue dash-dot line in Fig.~\ref{fig2}), which is $2.21\%$ higher than that in the laboratory frame. Owing to relativistic effects, the time coordinate is dilated from $t = 0.005$ in the laboratory frame to $t' = 0.004$ in the relative frame (see Fig.~\ref{fig2}).

\begin{figure*}[htbp]
    \centering
    \includegraphics[width=0.32\textwidth]{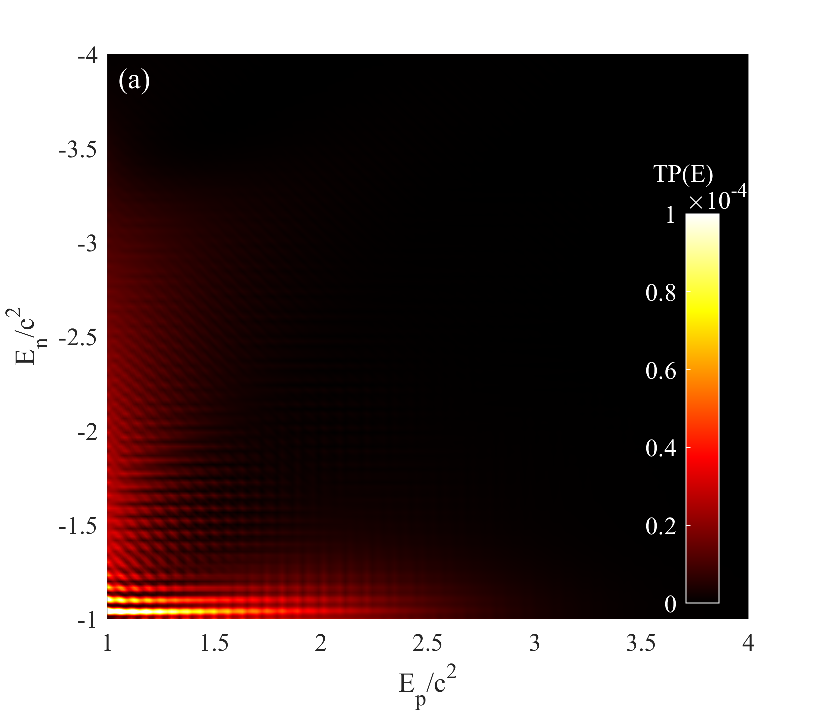}
    \includegraphics[width=0.32\textwidth]{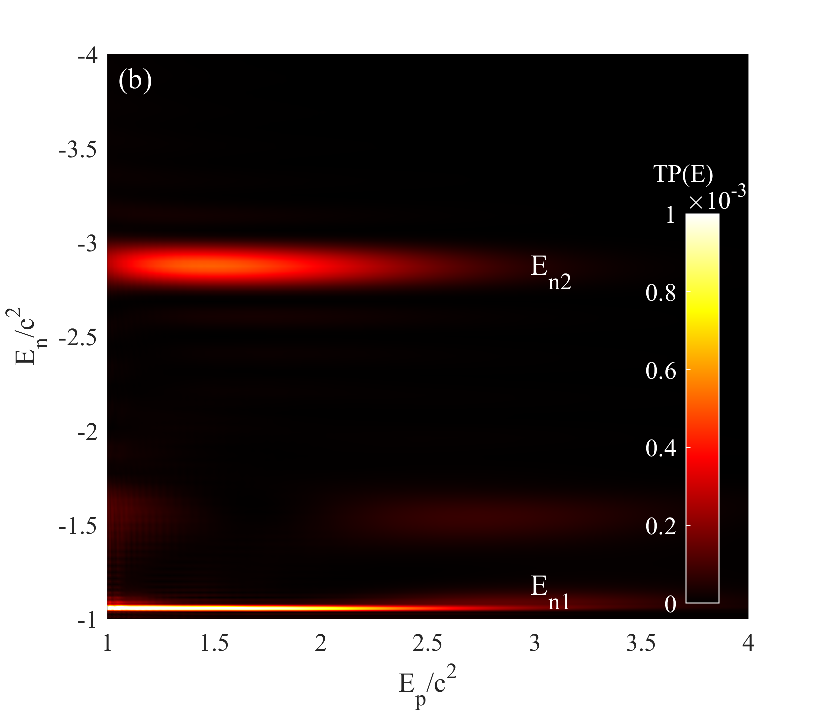}
    \includegraphics[width=0.32\textwidth]{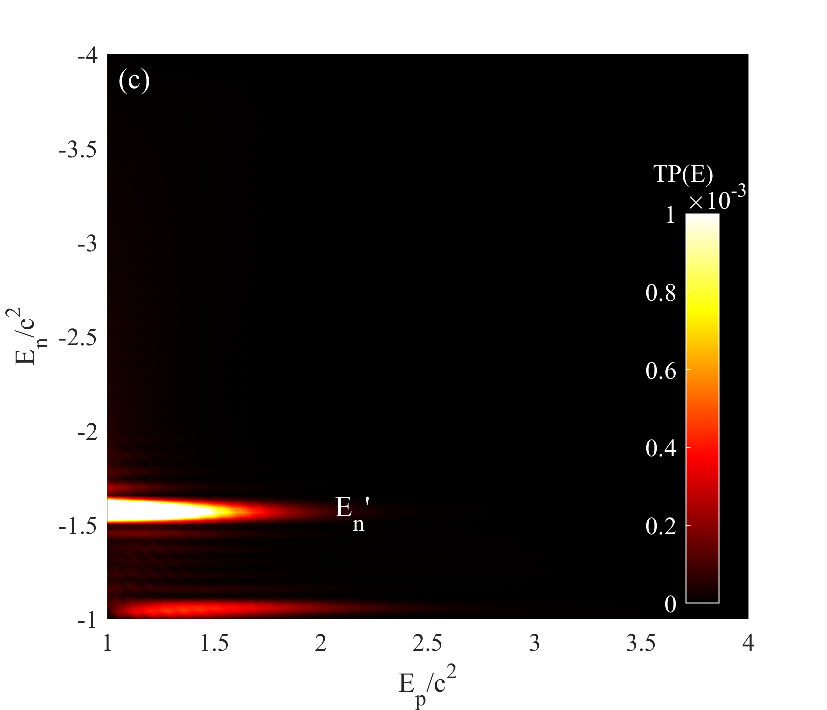}
    \caption{\label{fig3} 
    The tunneling probability of electron. The colorbar represents the particle tunneling probability.
    (a) The rest potential well in the laboratory frame, 
    (b) The relativistic moving potential well in the laboratory frame, 
    (c) The rest potential well in the relative frame.
    Note that the colorbar scale in (a) is one order of magnitude lower than the others. The parameters are the same as those in Fig.~\ref{fig2}.}
\end{figure*}

To gain further insight into electron-positron pair creation via relativistic effects, we introduce the tunneling probability of an electron from a negative-energy state $|E_n\rangle$ to a positive-energy state $|E_p\rangle$, as illustrated in Fig.~\ref{fig3}. In the figure, the horizontal axis represents positive energy, and the vertical axis represents negative energy. The Fig.~\ref{fig3} (a) displays the tunneling probability of an electron for a potential well at rest in the laboratory frame. We find that a small number of electrons in negative-energy states tunnel to positive-energy states. This tunneling arises from the instantaneous turn-on effect, as previously observed in Fig.~\ref{fig2} (black solid line). Fig.~\ref{fig3} (b) illustrates the tunneling probability for a relativistically moving potential well ($v_0 = 0.6c$) in the laboratory frame. We observe two tunneling regions, located at $E_1 = 1.0586c^2$ and $E_2 = 2.8854c^2$. For the potential well at rest in the relative frame, the tunneling probability exhibits a single region at $E' = 1.5934c^2$, as shown in Fig.~\ref{fig3}(c). Compared with Fig.~\ref{fig3}(a), the relativistic motion of the potential well induces electron-positron pair creation via the Schwinger effect. In addition, the tunneling probability in the relative frame is larger than that in the laboratory frame. Consequently, as observed in Fig.~\ref{fig2}, the rate of vacuum pair production in the relative frame is faster than that in the laboratory frame.

\begin{figure*}[htbp] 
    \centering
    \includegraphics[width=0.45\textwidth]{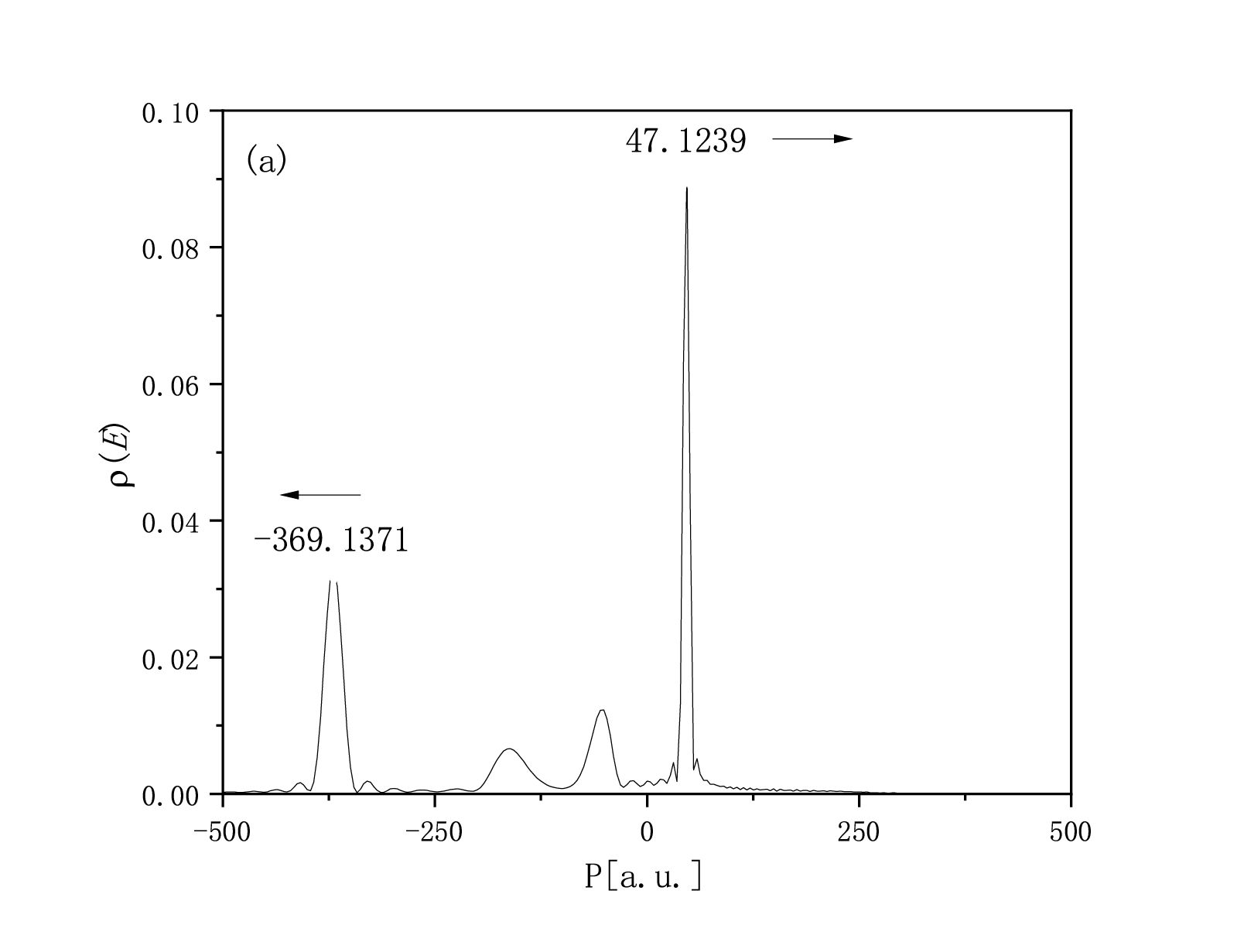}
    \includegraphics[width=0.45\textwidth]{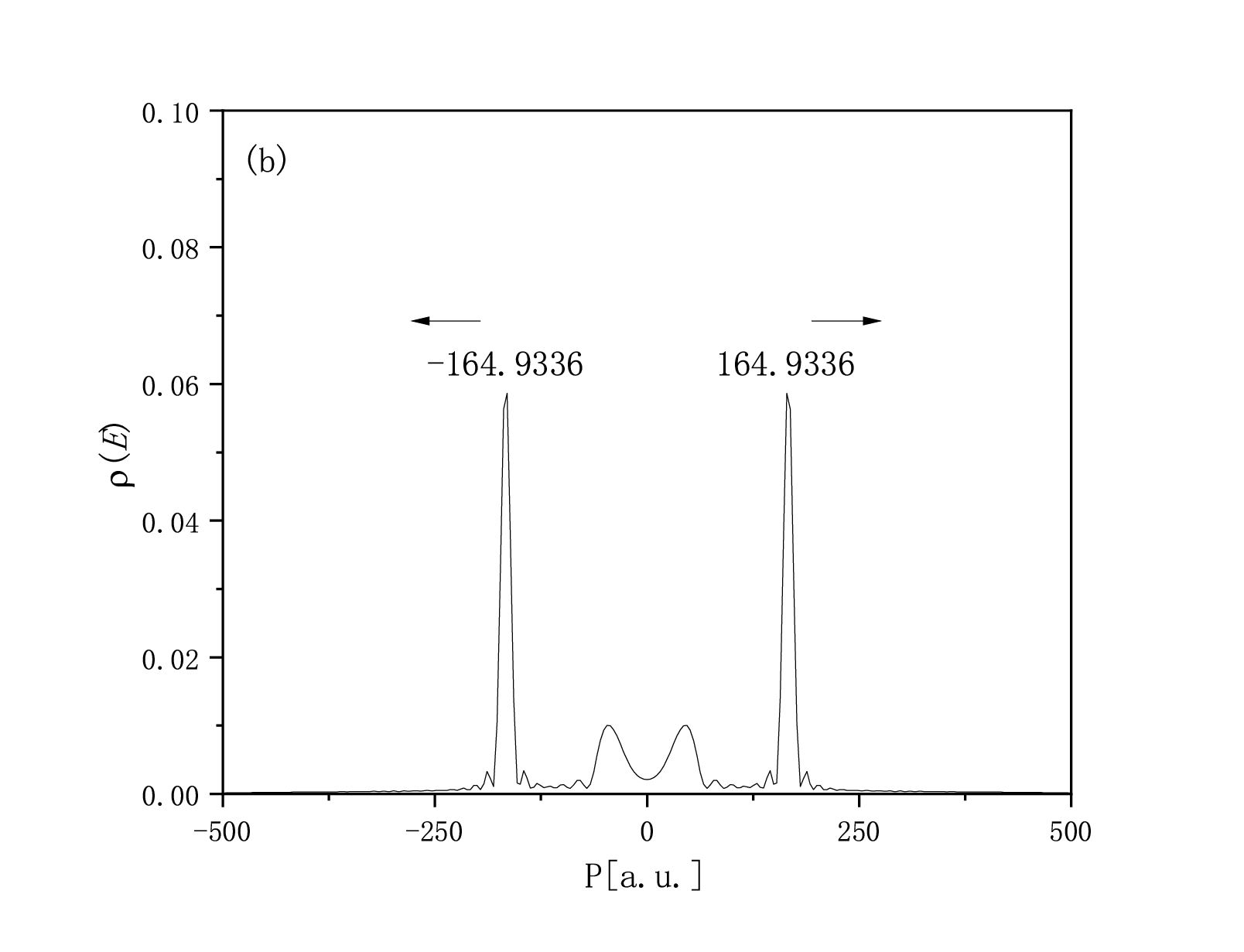}
    \caption{\label{fig4} The momentum spectra of the created positrons. (a) The relativistic moving potential well in the laboratory frame, (b) The rest potential well in the relative frame.  Arrows indicate the propagation directions of positrons. The parameters are the same as those in Fig.~\ref{fig2}.}
\end{figure*}

The corresponding momentum spectrum of the created positrons, shown in Fig.~\ref{fig4}, offers additional insight. We focus on the positron momentum spectrum rather than that of the electrons because the created electrons remain confined within the potential well. In contrast, positrons are promptly expelled from the interaction region by the repulsive potential, minimizing subsequent dynamic interference and allowing their energy distribution to effectively encode the characteristics of the creation process~\cite{LV2013}.
In the laboratory frame, two main momentum peaks are observed in Fig.~\ref{fig4}(a) at $P_1 = 47.1239$ and $P_2 = -369.1371$. Using the energy–momentum relation $E = \sqrt{c^2 p^2 + c^4}$, the corresponding energies of the created positrons are $E_1 = 1.0575c^2$ and $E_2 = 2.8740c^2$. In the relative frame, two main momentum peaks appear at $P' = \pm 164.9336$ in Fig.~\ref{fig4}(b), with an associated energy $E' = 1.5985c^2$. For the laboratory and relative frames, the tunneling regions are located at $E_1 = 1.0586c^2$ and $E_2 = 2.8854c^2$ [Fig.~\ref{fig3}(b)] and at $E' = 1.5934c^2$ [Fig.~\ref{fig3}(c)], respectively. The energies of the created positrons correspond to those of the negative-energy states from which the electrons tunnel. These results indicate that the tunneling probability $\left|U_{E_p,E_n}\right|^{2}$ circumvents the energy averaging issue inherent in integrating over the bound-state electron energy spectrum. This allows us to directly access the physical information of the bound states in electron calculations without requiring additional numerical computations for positrons, thereby providing an efficient new method for studying bound states in electron-positron pair creation.

\section{The Bound States Induced by the Relativistic Moving Potential Well}
\label{4}

As discussed in Sec.~\ref{3}, the number of bound states that couple to the negative-energy state depends on the reference frame. In the laboratory frame, two such resonant bound states are present, while only one appears in the relative frame. Another notable difference lies in the energy full width at half maximum (FWHM) of these bound states, which also varies between the two frames. To capture these phenomena in a general context, we introduce a precise multi-bound-state model. To this end, we consider a deep and sharp relativistic moving potential well with parameters $V = 2.8c^2$, $W = 0.7/c$, $D = 4/c$ (The equivalent electric field in SI units is $E=2.64\times 10^{18}\ \mathrm{V/m}$.), and reduce its speed from $v_0 = 0.6c$ to $v_0 = 0.55c$ in the laboratory frame.

\subsection{Bound State Combination Effect}
\label{4a}

\begin{figure*}[htbp]
    \centering
    \includegraphics[width=0.45\textwidth]{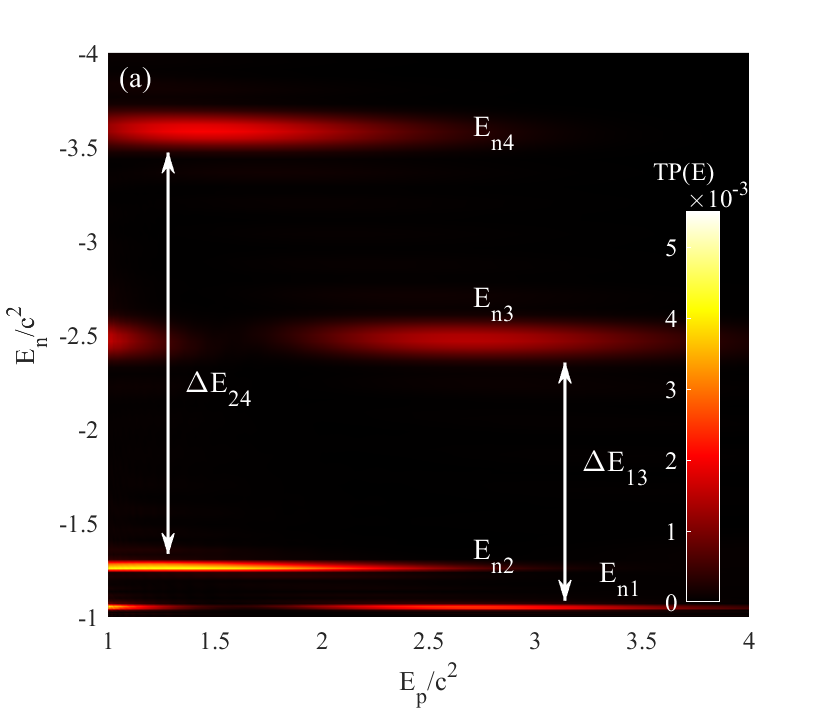}
    \includegraphics[width=0.45\textwidth]{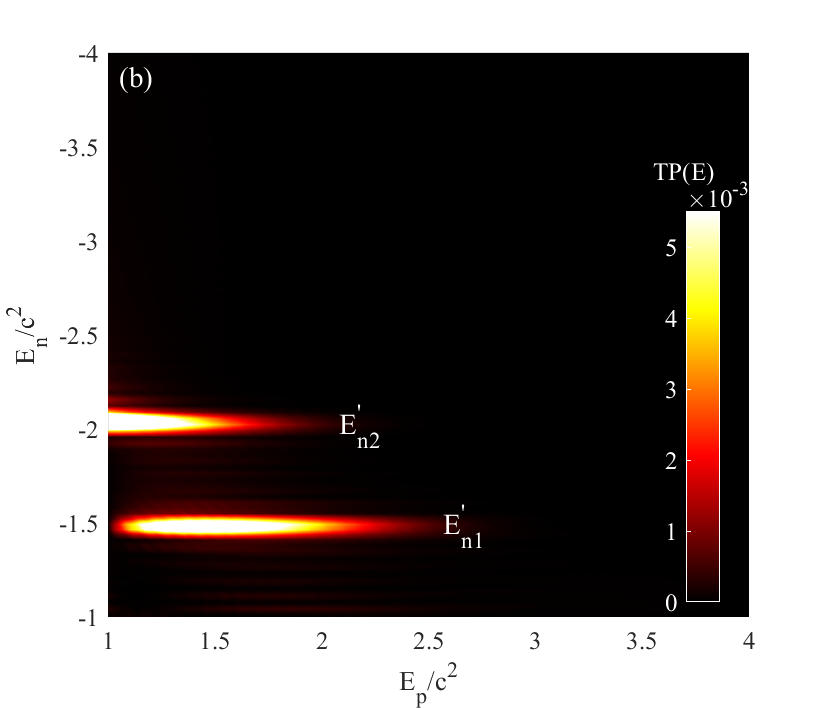}
    \caption{\label{fig5} The tunneling probability of electron. The colorbar represents the particle tunneling probability. (a) The relativistic moving potential well in the laboratory frame ($V=2.8c^2$, $W=0.7/c$, $D=4/c$, $v_0=0.55c$, $t=0.005$). (b) The rest potential well in the relative frame($V'=3.35c^2$, $W'=0.84/c$, $D'=4.79/c$, $v^{'}=0$, $t'=0.0042$).}
\end{figure*}

We analyzed the bound states resonating with the negative-energy state for the new potential well. Fig.~\ref{fig5} presents the tunneling probability of the electron for both reference frames. In the laboratory frame, four tunneling regions are observed in Fig.~\ref{fig5}(a), whereas in the relative frame, two tunneling regions appear in Fig.~\ref{fig5}(b). Accordingly, four bound states resonate with the negative-energy state in the laboratory frame, compared with two in the relative frame. The values of the four-dimensional energy--momentum tensor differ in the two reference frames, yet they are connected via the Lorentz transformation. Consequently, the number of bound states resonating with the negative-energy state is frame-dependent.

\begin{figure*}[htbp]
    \centering
    \includegraphics[width=0.45\textwidth]{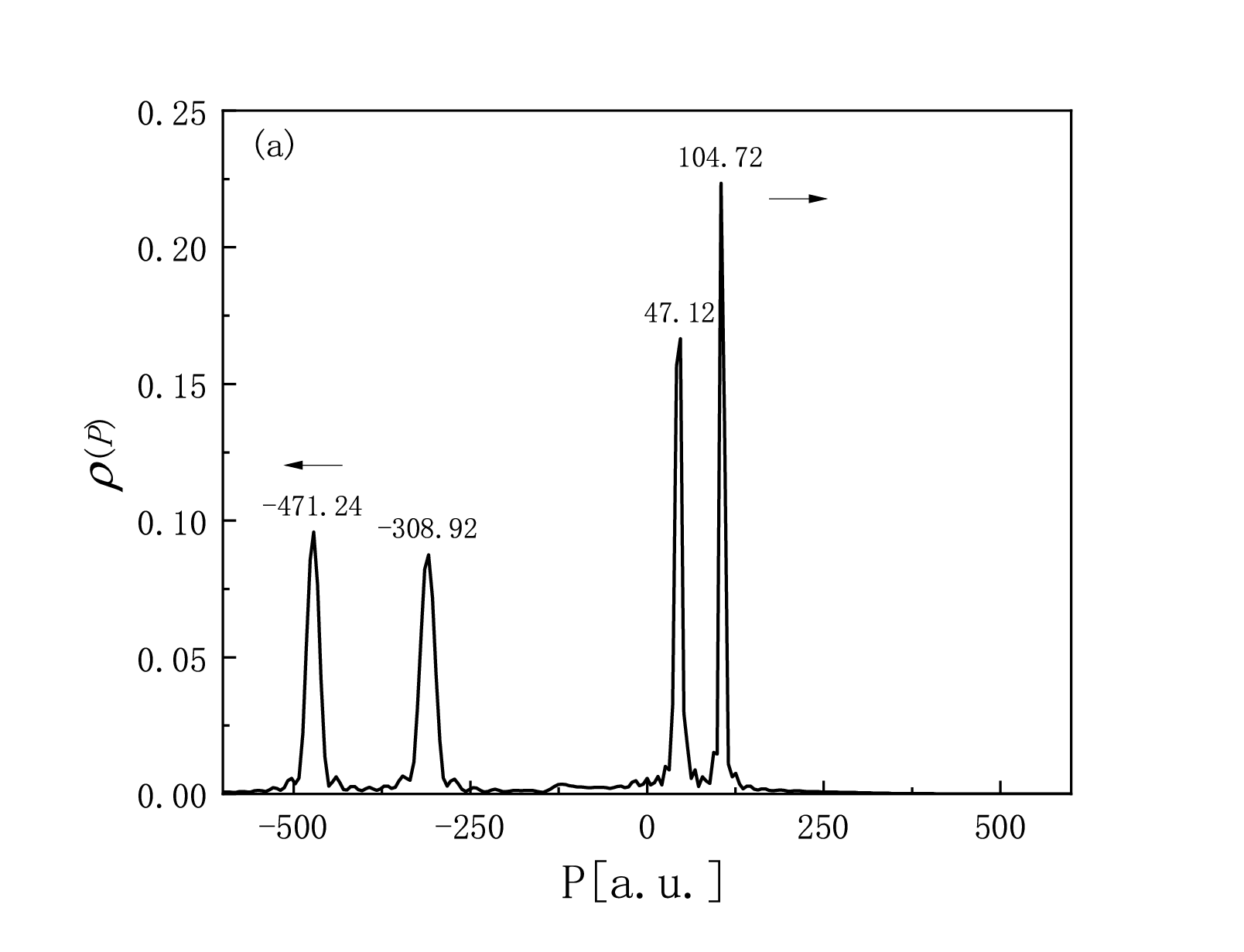}
    \includegraphics[width=0.45\textwidth]{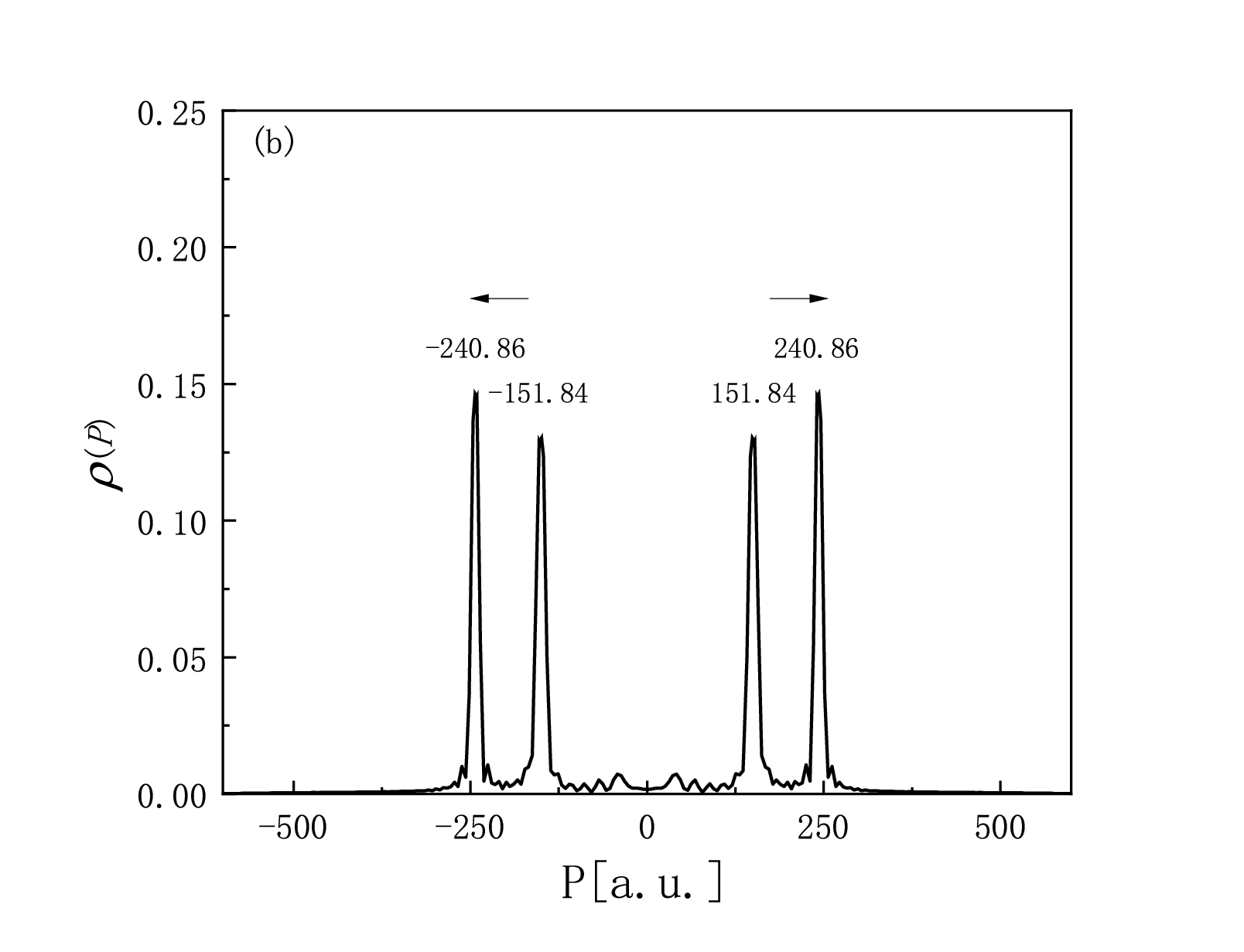}
    \caption{\label{fig6}The momentum distribution of the created positron. 
(a) The relativistic moving potential well in the laboratory frame. (b) The rest potential well in the relative frame. Arrows indicate the propagation directions of positrons. The parameters are the same as those in Fig.~\ref{fig5}.}
\end{figure*}

In four-dimensional spacetime, the energy and momentum of the created particles in the two reference frames obey the Lorentz transformation relations:
\begin{equation}
P_z' = \gamma\left(P_z - \frac{v_0}{c^2} E\right),
\label{eq13}
\end{equation}
\begin{equation}
E' = \gamma\left(E - v_0 P_z\right).
\label{eq14}
\end{equation}
Here, $P_z'$ and $E'$ ($P_z$ and $E$) represent the momentum and energy of the created particles in the relative (laboratory) frame. As discussed in Sec.~\ref{3}, the energy of the created positron matches that of the bound states. Notably, in contrast to the exact numerical evolution given by CQFT, IRFM enables a semi-analytical interpretation of the dynamics. Therefore, in the laboratory frame, the positron energies are $E_{n1} = -1.032c^2$, $E_{n2} = -1.259c^2$, $E_{n3} = -2.475c^2$ and $E_{n4} = -3.587c^2$, as shown in Fig.~\ref{fig5}(a). The corresponding momenta of the created positrons in the laboratory frame are $P_{z1} = 0.344c$, $P_{z2} = 0.764c$, $P_{z3} = -2.255c$ and $P_{z4} = -3.440c$, as shown in Fig.~\ref{fig6}(a). Based on Eqs.~\ref{eq13} and~\ref{eq14}, we calculate the transformation of the bound state energies between the laboratory and relative frames. In the relative frame, the bound state energies are $E_{n1}' = -1.462c^2$, $E_{n2}' = -2.010c^2$, $E_{n3}' = -1.478c^2$, and $E_{n4}' = -2.029c^2$, as listed in Table~\ref{table1}. The table also presents the numerical energies of the bound states obtained via CQFT. The numerical results are in good agreement with the analytical ones, with relative errors below $5\%$. The bound states $E_{n1}$ and $E_{n3}$ in the laboratory frame merge into $E_{n1}'$ in the relative frame, while $E_{n2}$ and $E_{n4}$ similarly merge into $E_{n2}'$. Thus, although the structure of the bound states differs between the two reference frames, the descriptions are connected via a Lorentz transformation.

The Lorentz transformation also governs the splitting between bound states, yielding energy separations of $\Delta E_{13} = |E_3 - E_1| = 1.443c^2$ and $\Delta E_{24} = |E_2 - E_4| = 2.328c^2$, as shown in Fig.~\ref{fig5}(a). The corresponding particle momenta in the relative frame are presented in Fig.~\ref{fig6}(b). Applying the Lorentz transformation, we obtain the bound-state splitting distances: $\Delta E_{13} = \gamma[v_0(P_{z3}' - P_{z1}')] = 1.459c^2$ and $\Delta E_{24} = \gamma[v_0(P_{z4}' - P_{z2}')] = 2.315c^2$. These analytical results are in good agreement with the numerical values above.

\begin{table}[htbp]
\centering
\caption{Analytical and numerical results for bound state energies}
\begin{tabular}{cc}
\hline

\textbf{Analytical(IRFM)} & \textbf{Numerical(CQFT)} \\
\hline 
$E_{n4}' = -2.029c^2$ & $E_{n2}' = -2.042c^2$ \\
$E_{n3}' = -1.478c^2$ & $E_{n1}' = -1.480c^2$ \\
$E_{n2}' = -2.010c^2$  & $E_{n2}' = -2.042c^2$ \\
$E_{n1}' = -1.462c^2$ & $E_{n1}' = -1.480c^2$\\

\hline
\end{tabular}
\label{table1}
\end{table}

\subsection{The Lifetime of Bound States}
\label{4b}

In this section, we demonstrate that relativistic effects also affect the energy FWHM of the bound states.We further derive how the FWHM of the bound-state energy transforms between reference frames by applying Lorentz transformations to the energy and momentum of the created particle pair. Transforming from the relative frame to the laboratory frame yields $|\Delta E| = \gamma ( |\Delta E'|$ $ + v_0 |\Delta P'_{z}| )$, and the inverse transformation can be verified accordingly. Here, $\Delta E$ is the energy FWHM in the laboratory frame, and $\Delta E'$ and $\Delta P'_{z}$ denote the energy FWHM and momentum FWHM, respectively, in the relative frame.

Within the framework of CQFT, we computed the electron tunneling probability [Fig.~\ref{fig5}(b)] and the corresponding momentum spectrum [Fig.~\ref{fig6}(b)] in the relative frame. The energy FWHMs of the bound states in the relative frame are $\Delta E_{n1}' = 0.0790c^2$ and $\Delta E_{n2}' = 0.0773c^2$, as listed in Table~\ref{table2}. The corresponding momentum FWHMs are $\Delta P'_{z1} = 0.1049c$ and $\Delta P'_{z2} = 0.0907c$. The energy FWHMs of the bound states in the laboratory frame are obtained analytically via the transformation relation $|\Delta E| = \gamma \left( |\Delta E'| + v_0 |\Delta P'_{z}| \right)$. For $\Delta E_{n1}' = 0.0790c^2$ in the relative frame, the corresponding FWHMs in the laboratory frame are $\Delta E_{n1} = 0.0255c^2$ and $\Delta E_{n3} = 0.1636c^2$. For $\Delta E_{n2}' = 0.0773c^2$, the corresponding FWHMs are $\Delta E_{n2} = 0.0328c^2$ and $\Delta E_{n4} = 0.1522c^2$. Meanwhile, the energy FWHMs of the bound states in the laboratory frame computed via CQFT are $\Delta E_{n1} = 0.0234c^2$, $\Delta E_{n2} = 0.0349c^2$, $\Delta E_{n3} = 0.1616c^2$, and $\Delta E_{n4} = 0.1543c^2$, as listed in Table~\ref{table2}. A comparison shows that the analytical results agree with the numerical ones to within numerical accuracy. Thus, the Lorentz transformation effectively facilitates the conversion between the two reference frames.

\begin{table}[htbp]
\centering
\caption{Comparison of energy broadening between reference frames}
\begin{tabular}{ccc}
\hline
\multirow{2}{*}{\textbf{Rel. frame}} & \multicolumn{2}{c}{\textbf{Lab frame}} \\
& \textbf{Numerical(CQFT)} & \textbf{Analytical(IRFM)} \\
\hline
$\Delta E_{2}' = 0.0773c^2$ & $\Delta E_{4} = 0.1522c^2$ & $\Delta E_{4} = 0.1543c^2$ \\
$\Delta E_{1}' = 0.0790c^2$ & $\Delta E_{3} = 0.1636c^2$ & $\Delta E_{3} = 0.1616c^2$ \\
$\Delta E_{2}' = 0.0773c^2$ & $\Delta E_{2} = 0.0328c^2$ & $\Delta E_{2} = 0.0349c^2$ \\
$\Delta E_{1}' = 0.0790c^2$ & $\Delta E_{1} = 0.0255c^2$ & $\Delta E_{1} = 0.0234c^2$ \\
\hline
\end{tabular}
\label{table2}
\end{table}

Applying the uncertainty principle $\Delta E \times \Delta t \sim \hbar$ provides insight into the relevant time scales: a narrow energy FWHM indicates a long lifetime, whereas a broad FWHM indicates a short one. Consequently, the measured energy FWHM characterizes the lifetime of the bound state. In the relative frame, the energy FWHMs $\Delta E_{n1}'$ and $\Delta E_{n2}'$ are comparable (Table~\ref{table2}), suggesting that the ground state $E_{n1}'$ and the first excited state $E_{n2}'$ have similar lifetimes. In the laboratory frame, however, $\Delta E_{n3}$ is approximately 6.5 times larger than $\Delta E_{n1}$, and $\Delta E_{n4}$ is about 3.5 times larger than $\Delta E_{n2}$ (Table~\ref{table2}). Consequently, the lifetimes of the high-energy states ($E_{n3}$ and $E_{n4}$) are significantly shorter than those of the low-energy states ($E_{n1}$ and $E_{n2}$). As a result, the contribution of high-energy states to electron-positron pair creation is much smaller~\cite{SU2020}. 

From a relativistic perspective, this effect can be understood in terms of the invariant interval $S^2 = S'^2$, a fundamental property of spacetime analogous to length contraction. As detailed in Table~\ref{table2}, the energy FWHMs $\Delta E_{n1}$ and $\Delta E_{n3}$ in the laboratory frame originate from $\Delta E_{n1}'$ in the relative frame, while $\Delta E_{n2}$ and $\Delta E_{n4}$ originate from $\Delta E_{n2}'$. This mapping is consistent with the bound state combination effect discussed in Sec.~\ref{4a}, suggesting that such combination behavior reflects a property of spacetime itself. Notably, while the number and lifetime of bound states vary with reference frame, the particle number remains invariant, independent of the frame.

It is noted that the combination of bound states when transforming from the laboratory frame to the relative frame can equivalently be viewed as a splitting of bound states when transforming in the opposite direction. This splitting effect is different from the energy level splitting typically caused by the electromagnetic field, such as the Zeeman effect. It is a unique phenomenon resulting from the characteristics of spacetime. In the relative frame, the physical system exhibits spatial reflection symmetry. For the quasi-bound states in the potential well, particle currents flowing in the positive and negative directions contribute equally to the kinetic energy, as shown in Fig.~\ref{fig6}(b). This mirror symmetry leads to the observed degeneracy of energy levels in this reference frame. In the laboratory frame, the original spatial symmetry of the system is broken by the Lorentz boost. According to the energy–momentum transformation $E' = \gamma\left(E - v_0 P_z\right)$, particles moving in different directions experience distinct Doppler-like shifts: forward-moving particles gain energy due to velocity addition, while backward-moving particles lose energy. Therefore, the observed merging or splitting of energy levels arises from the fact that the kinetic energy term is not invariant across reference frames, leading to a significant shift in the eigenvalues of the initially degenerate states. Two symmetric states that are indistinguishable in the relative frame are "pulled apart" in the laboratory frame due to their momentum difference, manifesting as level splitting. Conversely, when multiple levels are asymmetrically translated and overlap on the energy axis, they appear merged in the spectrum. This phenomenon indicates that the lowered threshold for vacuum electron-positron pair creation stems not merely from an increased potential depth, but more fundamentally from a relativistic phase-space reconfiguration that optimizes the vacuum excitation pathway in the energy dimension.

\section{Conclusion}
\label{5}
In this work, we have extended the CQFT and IRFM frameworks to investigate electron-positron pair creation induced by a relativistic moving potential well. By defining the electron tunneling probability within CQFT, we observe that the same potential well exhibits distinct resonant bound-state characteristics across reference frames. From this probability, we extract the bound-state energy, energy FWHM, and lifetime. This approach enables direct access to bound-state information within CQFT without additional numerical computations, offering an efficient new tool for studying bound states in electron-positron pair creation.

An intriguing phenomenon is observed: the evolution of bound states within the same potential well is frame-dependent. Notably, the resulting splitting effect differs from conventional electromagnetic-field-induced level splitting. By developing the IRFM, we show that this discrepancy originates from the Lorentz transformation of the four-momentum between reference frames.

Furthermore, our findings show that these relativistic effects enable a reduction in the threshold electric field for electron-positron pair creation. By either accelerating an atomic system or shaping a subcritical electric field to enter the supercritical regime, the effective threshold can be substantially lowered. Experimentally, heavy ions can be accelerated to relativistic energies, and their Coulomb fields act as relativistically moving potential wells in the laboratory frame, thereby producing electron-positron pairs. Meanwhile, this offers promising prospects for experimental implementation at next-generation high-intensity laser facilities.

\section*{Acknowledgements}
This work has been supported by the Strategic Priority Research Program of the Chinese Academy of Sciences (Grants No. XDA25051000), National Natural Science Foundation of China (Grants No. 12447120 and No. 11974419), Natural Science Foundation of Henan (Grant No. 252300423526).

\section*{Data Availability Statement}
This manuscript has no associated data. [Authors' comment: Data sharing not applicable to this article as no datasets were generated or analysed during the current study.]

\section*{Code Availability Statement}
This manuscript has no associated code/software. [Authors' comment: Code/Software sharing not applicable to this article as no code/software was generated or analysed during the current study.]



\end{document}